# Kondo-effect protected topological surface states on Sb(111)

Limin She[1,2], Yinghui Yu[1,*], and Gengyu Cao[1,*]

[1]*State Key Laboratory of Magnetic Resonance and Atomic and Molecular Physics, Wuhan Institute of Physics and Mathematics, Chinese Academy of Sciences, Wuhan 430071, PR China*

[2]*University of Chinese Academy of Sciences, Beijing 100049, PR China*

*Author to whom all correspondence should be addressed.
E-mail address: yhyu@wipm.ac.cn and gycao@wipm.ac.cn

## Abstract

Local density of states and quasiparticle interference phenomena of Co/Sb(111) were investigated by scanning tunneling spectroscopy. A sharp peak observed near the Fermi energy is interpreted as a fingerprint of the conventional Kondo resonance with the Kondo temperature of $205 \pm 23$ K. Moreover, by identifying the interference wavevectors, only the scattering channels relating to backscattering confinements are observed for surfaces with and without the Co deposition. It reveals that the Kondo effect fully screens the magnetic impurities and thereby suppresses the backscattering of the topological surface state.

## Introduction

Topological insulators (TIs) as a new class of spin-orbital coupling materials are characterized by spin-splitting surface states with linear energy dispersions. The novel topological surface states (TSS) following massless Dirac equation are protected by time-reversal (TR) symmetry and are insensitive to spin-independent backscattering [1-4]. These remarkable properties ensure TIs as a promise matter for applications in fields of spin electronic devices and quantum information processing [5-6]. Many experiments have already demonstrated that the chiral spin texture and TR symmetry prohibit the electrons of TSS backscattering from nonmagnetic defects [7-11]. Meanwhile, the coupling of Dirac fermions to local magnetic moments have attracted considerable interests in fields of condensed-matter physics, since the TR symmetry of TSS may be locally broken by magnetic impurities [12-14]. Up to now, a lot of experimental work has focused on this aspect [15-23] and the stability of the TSS against local magnetic perturbations is still in controversy [17-23]. For example, Fe bulk-doped $Bi_2Te_3$ presents a new scattering channel in quasiparticle interference (QPI) phenomena due to the broken TR symmetry [15] and Fe-deposited $Bi_2Se_3$ systematically modifies the topological spin structure, leading to TR breaking [17]. Nevertheless, cases of magnetic impurities such as manganese, iron, cobalt, and gadolinium adsorbed on TIs indicate that the surface states are remarkably insensitive to magnetic impurities that are not capable of breaking TR symmetry [19-23].

Furthermore, a few theoretical calculations predict that a Kondo effect would take place as long as the Fermi level do not locate exactly at the Dirac point, leading to spin-polarized electronic clouds which screens the magnetic moments of local impurities and thus suppresses the backscattering of the TSS [24-29]. A most recent experiment [30] carried out on the clean surface of Kondo insulator $SmB_6$ implies that the Kondo interaction commonly observed on noble metals [31-34] is compatible with the TSS [35]. One interesting question addressed is then whether the conventional Kondo effect can be observed and exactly protect the TSS from magnetic scattering when magnetic adsorbates couple to the TSS.

Angle-resolved photoemission spectroscopy (APRES) and scanning tunneling microscopy (STM) results have already clarified that the surface states of Sb(111) was protected by the TR symmetry [9, 36-37]. In this work, the local density of states (LDOS) of Co atoms adsorbed on Sb(111) were investigated by scanning tunneling spectroscopy (STS). A sharp peak locating near the Fermi energy arises from the Kondo resonance that is assigned to fully screening the localized spin of Co impurities by the surrounding Kondo electron cloud. Furthermore, the quasiparticle scattering phenomena around Co atoms are recorded by energy resolved differential conductance (*dI/dV*) mapping. By carefully identifying the QPI patterns, we clarify the scattering channels corresponding to forbidden backscattering for surfaces with and without the Co adatoms. These results conclude that the TR symmetry of the Dirac fermions is protected from the magnetic scattering by the Kondo effect.

## Experiments

The experiments were carried out in an ultrahigh vacuum low temperature STM (Unisoku, Japan) with a base pressure better than $1 \times 10^{-8}$ Pa. The Sb(111) substrate was *in situ* cleaned by repeated cycles of $Ar^+$ sputtering and subsequently annealing to about 620 K. The surface quality was checked by STM until the images showed no distinct traces of contaminants. After the substrate temperature was recovered to room temperature, Co atoms were deposited from an e-beam evaporator equipped with a high purity Co rod (2.0 mm diameter, 99.995% purity). The coverage was confirmed by measuring Co islands grown on Cu(111) [38]. For better topography and energy resolution, here all STM measurements were performed at liquid helium temperature (about 4.5 K) with Pt-Ir tips. The *dI/dV* spectra were recorded by a lock-in technique with a closed feedback loop. Sinusoidal modulation voltage with the value of 4-6 mV was applied at a frequency of about 1660 Hz.

## Results and discussion

Figure 1(a) shows a typical STM topography imaged on clean Sb(111). The atomically resolved STM image (inset of Fig. 1(a)) exhibits the hexagonal lattice structure where the lattice orientation is marked by arrows. To further understand the quantum interaction between the TSS and magnetic impurities, a small amount of Co atoms are deposited on Sb(111). Figure 1(b) shows a topographic image of Sb(111) with about 0.01 monolayer (ML) Co adsorption. These Co atoms form bright protrusions on the surface with typical sizes ranging from 0.9 to 4 nm and are identified as a single adsorbed atom or clusters. Interestingly, some Co atoms diffuse downwards into the subsurface and exhibit an extended feature with a weak brightness as observed in Fig. 1(b). Here we identify the smallest bright protrusion to be a single Co atom as marked by a blue square (Fig. 1(b)). The line profile shown in Fig. 1(c) reveals that the single Co atom with the quite circular shape possesses a height of ~1 Å and a size (full width at half maximum) of ~0.9 Å, similar to the case of a single Co atom adsorbed on Cu(100) that shows a height of ~1.1 Å and a size of ~0.8 Å [32].

To understand the LDOS evolution, STS measurements are performed on surfaces with and without the Co deposition (Fig. 1d). The $dI/dV$ curves exhibits very similar feature of LDOS for both surfaces. There are three obvious peaks in the $dI/dV$ spectra of clean Sb(111) locating at the energies of ~-230, ~-120 and ~230 meV, respectively. As previously reported [36-37, 39], the surface state bands of Sb(111) can be viewed as a distorted Dirac cone. The electron pocket of the cone bands extends upwards to the conduction band, but the hole band bends at the energies of about -120 and 230 meV and ultimately decays into the valance bands at the fixed orientation. In Fig. 1(d), the Dirac point of the cone band structure is pointed by a red arrow at the energy of about 230 meV below the Fermi level $E_F$. The peak features at about -120 and 230 mV just correspond to the bending position of the hole band along the $\bar{\Gamma} - \bar{K}$ and $\bar{\Gamma} - \bar{M}$ directions, respectively [37, 39]. The blue curve in Fig. 1(d) shows the LDOS recorded on Sb(111) with the ~0.01-ML Co deposition where three peaks still exist. Interestingly, we notice that all of these peaks shift towards the side of high bias voltages by about 25 mV in comparison with the corresponding features of the clean substrate. That is to say, the whole surface states of Sb(111) shifts by ~25 meV towards the vacuum energy level after Co atoms are deposited. In fact, similar characteristic has been observed elsewhere [16-17, 19-20, 23]. Considering relatively rich valence electrons of antimony, it is naturally expected that the charge transfer occurs from Sb to neighboring Co atoms. The surface charge redistribution should be responsible for the upward energy shift observed in Fig. 1(d).

In order to investigate the quantum interaction between TSS and magnetic impurities, we further probe the $dI/dV$ spectra by placing tips at the center of a single Co atom (Fig. 2). The spectrum presents a sharp peak near the Fermi level. Since the width of bare d resonances are universal on the order of 100 meV [40-41] that is much larger than our observation and also the peak position are very close to $E_F$, the possibility of a d band resonance is excluded. This sharp feature is very similar to the previous observation on noble metal surfaces that is assigned as a fingerprint of Kondo resonance [31-33]. Therefore, here we believe that the sharp peak near $E_F$ should be attributed to the Kondo effect resulted from the full screening of local magnetic impurities by surrounding spin-polarized electron clouds.

The observed line shape of Kondo resonance can be understood by the conventional Fano model [42] that has been successfully applied to explain the interaction between electrons and localized spins for cases of magnetic impurities adsorbed on noble metals and the topological Kondo insulator $SmB_6$ [30-33]. The $dI/dV$ spectra near $E_F$ can be fitted to a Fano function [31-33, 42]

$$\frac{dI}{dV} = A + B\frac{(\varepsilon+q)^2}{1+\varepsilon^2}$$

with $\varepsilon = \frac{eV - \Delta E}{\Gamma/2}$ where A is the background $dI/dV$ signal, B is the amplitude coefficient, $q$ is the Fano line shape factor, $\Delta E$ is the energy of the Kondo resonance, and $\Gamma$ is the full width of the resonance. Via fitting the $dI/dV$ spectra recorded by using different tips to the Fano function, we acquire the average values of the fitting parameter $\Delta E = -7.3 \pm 2$ meV, $q = -9.8 \pm 4.3$ and $\Gamma = 50 \pm 6$ meV. Moreover, another important signature of the fitted Fano resonance is that the full width $\Gamma$ relates to the Kondo temperature $T_k$ with $\Gamma = 2\sqrt{(\pi k_B T)^2 + 2(k_B T_K)^2}$ that is deduced from Fermi liquid theory [33], where $T$ is the measurement temperature and $k_B$ is Boltzmann constant. Using this equation, it is easy to acquire the Kondo temperature $T_k = 205 \pm 23$ K. At temperatures below $T_k$, the local spin of a Co adatom is fully confined by the Kondo screening electron cloud of Sb(111).

As recent theoretical studies [24-29], Kondo screening electron cloud and the magnetic impurities can combine into a many-body spin singlet state as a result of the Kondo interaction and may suppress the backscattering of the TSS. So we further investigate the quantum interference fingerprints of TSS around Co impurities. The energy-resolved $dI/dV$ maps on the clean substrate surface are taken at the area highlighted by a rectangle in Fig. 1(a). Two typical $dI/dV$ maps are shown in Figs. 3(a) and 3(b), respectively. The line-shaped patterns along the straight step edge are similar to those commonly observed on noble metal surfaces [43] and are ascribed to the electronic standing waves resulting from QPI between electrons with different momenta. In Fig. 3(a), there are two types of alternating patterns with different oscillation wavelengths, and the corresponding fast Fourier transform (FFT) (inset of Fig. 3(a)) reveals two different quantized wavevectors, $q_1$ and $q_2$, for both oscillation patterns (the inner and outer spots correspond to $q_1$ and $q_2$, respectively.). In Fig. 3(b), however, only one type of oscillation patterns is observed and correspondingly one set of the wavevector is displayed in the FFT image (inset of Fig. 3(b)). As previously studied [9], both types of oscillation patterns in Fig. 3(a) are identified as originating from $q_1$ with scattering between the adjacent hole pockets and $q_2$ with interfering between the centered electron and corner hole pockets with oppositely oriented momenta and parallel spins. Whereas, at the energies below about -120 meV, the hole pockets of distorted cone bands overlap each other into a hexagonal shape, leading to vanishing of $q_1$ [37]. Thus the observed wavevector in the inset of Fig. 3(b) corresponds to $q_2$ and thereby only one type of oscillation patterns is imaged in Fig. 3(b).

The energy-resolved *dI/dV* mapping was further carried out at the surface containing Co adatoms. Figures 3(c) and 3(d) are two typical *dI/dV* maps and show rich interference fingerprints that are aroused by the surface state scattering around the Co defects. Oscillation patterns with anisotropic shapes emerge at the vicinity of those adsorbed Co atoms. The spatial periodicity of the QPI patterns increases rapidly along with the bias voltages approaching to Dirac point, whereas the corresponding scattering wavevectors reduces at the same time as shown in the FFT images (insets of Figs. 3(c) and 3(d)). Two sets of extended bright spots reside around $\bar{\Gamma}$ in the reciprocal space as shown in the inset of Fig. 3(c). The regions with high intensities are all oriented along the $\bar{\Gamma}-\bar{M}$ directions, while the intensity along the $\bar{\Gamma}-\bar{K}$ directions vanishes. This feature should be ascribed to the hole bands of Sb(111) decaying into the bulk valence band in the $\bar{\Gamma}-\bar{K}$ direction at the energy of about -120 meV [9, 37, 39]. With the energy decreasing, the inner bright regions quickly shrink towards $\bar{\Gamma}$, but the outer ones are still visible at the $\bar{\Gamma}-\bar{M}$ direction. At the energies below about -120 meV, for example -140 meV (inset of Fig. 3(d)), the outer spots extend into a ringlike shape centering at $\bar{\Gamma}$.

The scattering processes can be understood by analyzing the possible origin of QPI wavevectors $\bar{q}$ in terms of the constant energy contour (CEC). As noted above, q mainly emerges along the $\bar{\Gamma}-\bar{M}$ direction. Based on the CEC of Sb(111), we propose all possible scattering channels ($q_1$-$q_4$) relating to QPI patterns along the $\bar{\Gamma}-\bar{M}$ direction as shown in Figs. 4(a) and 4(b). The vectors $q_1$ and $q_2$ correspond to those on clean Sb(111) mentioned above, while the wavevector $q_3$ ($q_4$) correspond to the QPI between electron (hole) pockets with oppositely oriented momenta and spin along the $\bar{\Gamma}-\bar{M}$ direction. Concerning the topological protection nature, the vectors $q_3$ and $q_4$ relate to the backscattering of Dirac fermions that are forbidden by the TR symmetry. The two scattering channels must be considered, since the TR invariant of TSS may be broken by the Co adsorbates.

By measuring the interference patterns and corresponding FFT at various biases, we obtained the energy dispersions of QPI phenomena along the $\bar{\Gamma}-\bar{M}$ direction. In order to identify the scattering channels, the ARPES and STM data previously published are used to calculate the interference vectors as a function of the quasiparticle energy [9, 36-37]. The dispersions of the four possible scattering processes are shown in Fig. 4 (c) as gray solid curves ($q_1$ is obtained from previous STM data and $q_2$-$q_4$ are calculated from ARPES data.). The triangular and rectangular points correspond to the QPI wavenumbers that are acquired from the Sb(111) surface with and without Co impurities, respectively. It is obvious that the rectangular points well match the calculated $q_1$ and $q_2$ that are protected by TR symmetry [9]. However, the experimental values characterized by triangles do not totally match the calculated $q_1$ and $q_2$, and both dispersions wholly shift by about 25 meV towards the vacuum energy level in comparison with those of clean Sb(111). This dissimilarity is identical to the observed energy shift in Fig. 1(d) which is caused by the surface charge transfer from Sb to neighboring Co atoms. Considering the energy shift, accordingly, only the scattering channels protected by the TR symmetry are observed on the surfaces with and without Co adatoms. That is to say, the backscattering of the TSS is still forbidden, although the magnetic moment of Co is expected to break the TR symmetry.

A straightforward question is then whether the conventional Kondo effect observed in Fig. 2 influences the QPI between Dirac fermions and magnetic impurities. According to recent theoretical calculations [26, 28, 29], Kondo screening electronic cloud is formed around magnetic adsorbates that are fully Kondo screened for Dirac point of TSS away from $E_F$ and temperatures much below $T_k$. For Sb(111), the Dirac point locates at the energy of about 230 meV below $E_F$ as mentioned above. Thus the local magnetic moment of Co is fully screened by the Kondo electron clouds at the experimental temperature (about 4.5 K) that is much lower than $T_k$ ($205 \pm 23$ K). In this case, the Co adsorbates and the Kondo screening clouds combine into a many-body spin singlet state and serve as a nonmagnetic potential scatterer. Consequently, the QPI fingerprints observed in Figs. 3(c) and 3(d) originate from the TSS' electrons scattering from the nonmagnetic Kondo state. In other words, Kondo screening protects the TSS from magnetic scattering, and the TR symmetry of the TSS sustains on Co deposited Sb(111) as revealed in Fig. 4(c).

## Conclusions

By STS, the conventional Kondo effect with a Kondo temperature of $205 \pm 23$ K is observed on a single Co adatom. Moreover, the QPI phenomena of the TSS around Co impurities are investigated by Fourier transform STM. No hints of broken TR symmetry are observed, although the surface state presents a blue energy shift of ~25 meV due to the surface charge transfer. It is concluded that the Kondo screening can protect the TSS from magnetic scattering. These results provide a pathway to further understanding the interaction between the topological nontrivial states and magnetic adsorbates.

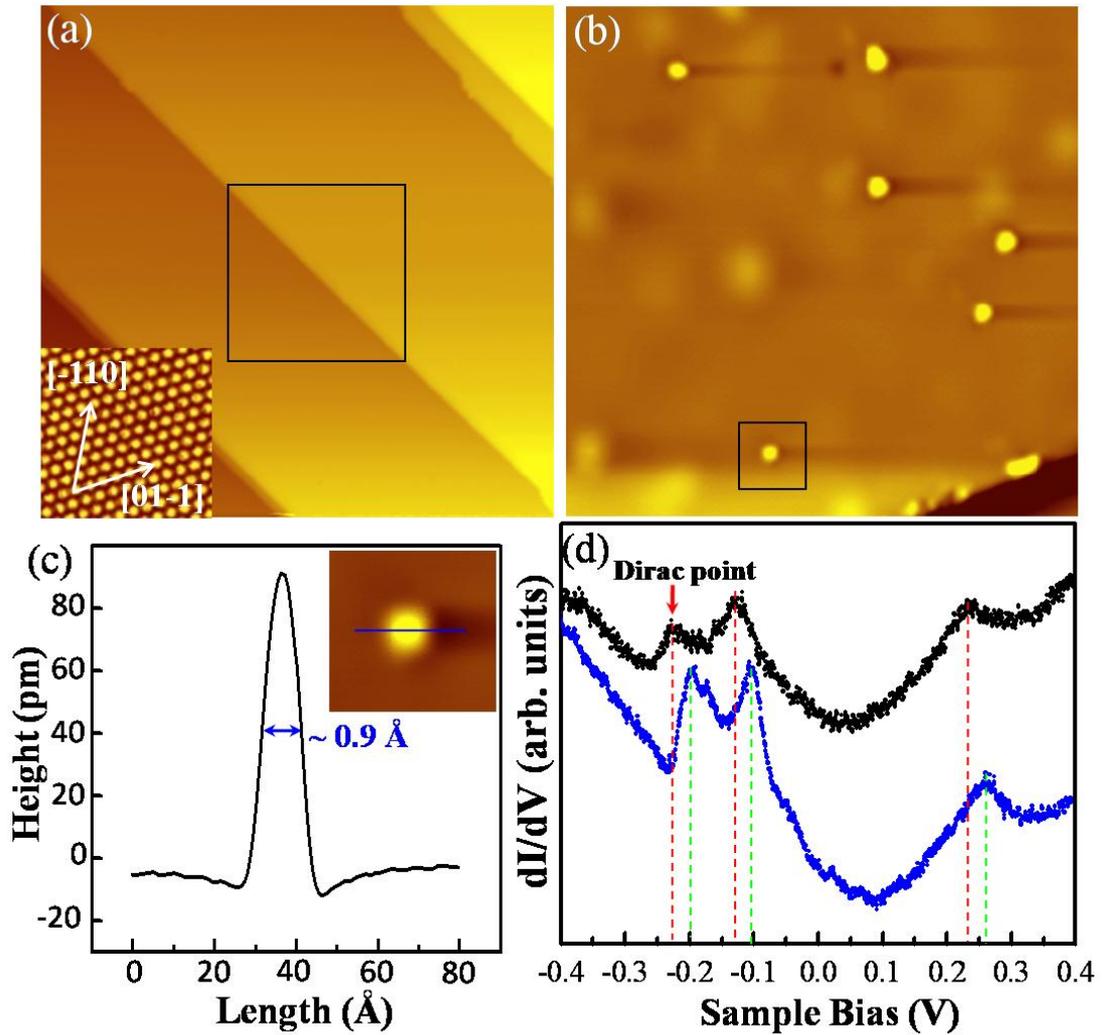

FIG. 1 (a) A typical STM topography of Sb(111) (113 nm × 113 nm; U = 0.4 V; I = 30 pA). The inset shows an atomically resolved STM image where the lattice orientations are indicated by arrows. (b) A STM topography of Sb(111) with the 0.01-ML Co deposition (40 nm × 40 nm, U = 1.5 V, I = 20 pA). (c) Line profile taken along the line in the inset that is a zoomed-in STM image of the region indicated by a rectangle in (b). (d) *dI/dV* spectra recorded on Sb(111) with (blue) and without (black) the Co deposition. (set point: U = 0.4 V, I = 100pA).

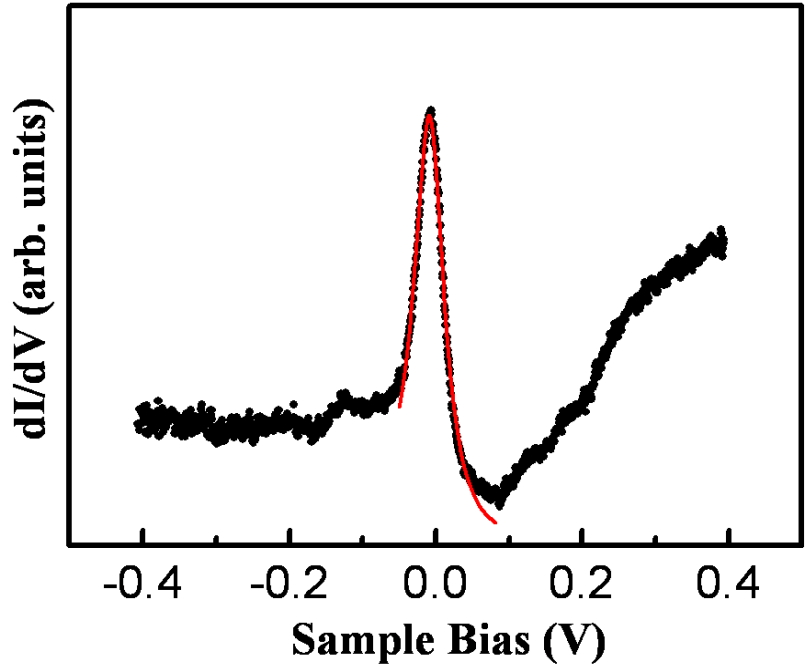

FIG. 2 *dI/dV* spectra recorded at the center of single Co atoms (set point: U = 0.4 V, I = 10 pA). The sharp peak is fitted by Fano function (red curve). The fitting parameters are $\Delta E = -7.3 \pm 2$ meV, $q = -9.8 \pm 4.3$ and $\Gamma = 50 \pm 6$ meV.

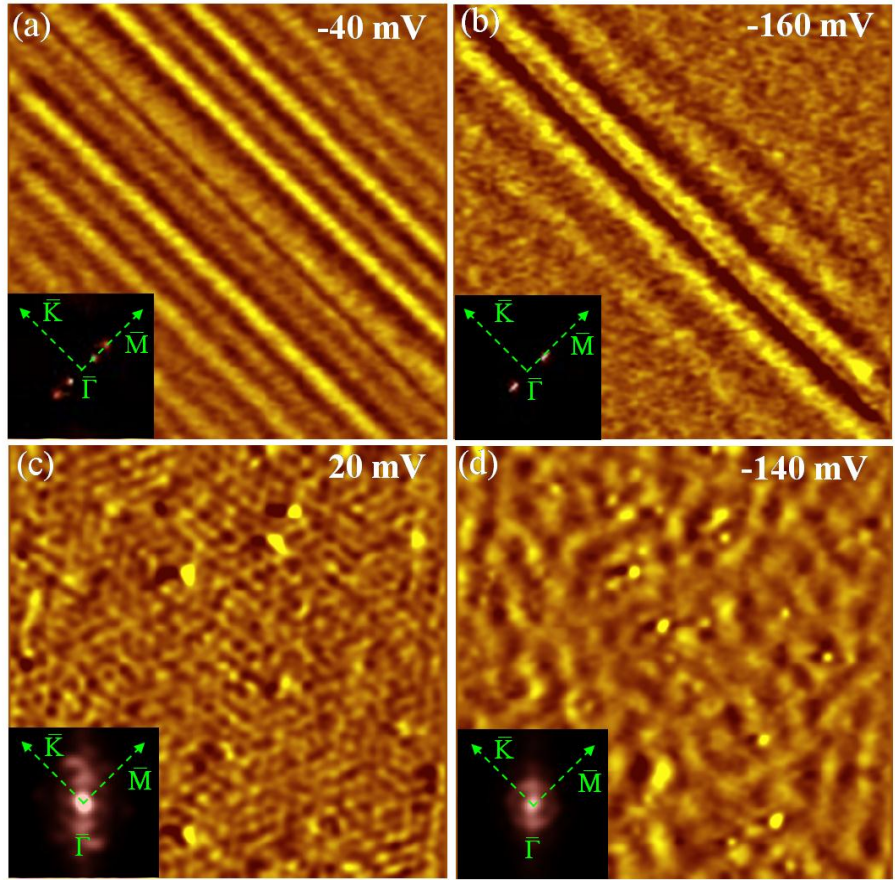

FIG. 3 (a) and (b) Energy-resolved *dI/dV* maps of the region indicated by a rectangle in Figure 1(a). (51.5 nm × 51.5 nm, setpoint: I = 80 pA). (c) and (d) Energy-resolved *dI/dV* maps recorded on ~0.01-ML Co deposited Sb(111). (100 nm × 100 nm, set point: I = 100 pA). The corresponding FFT power spectra are shown in the insets where the reciprocal directions are marked by arrows.

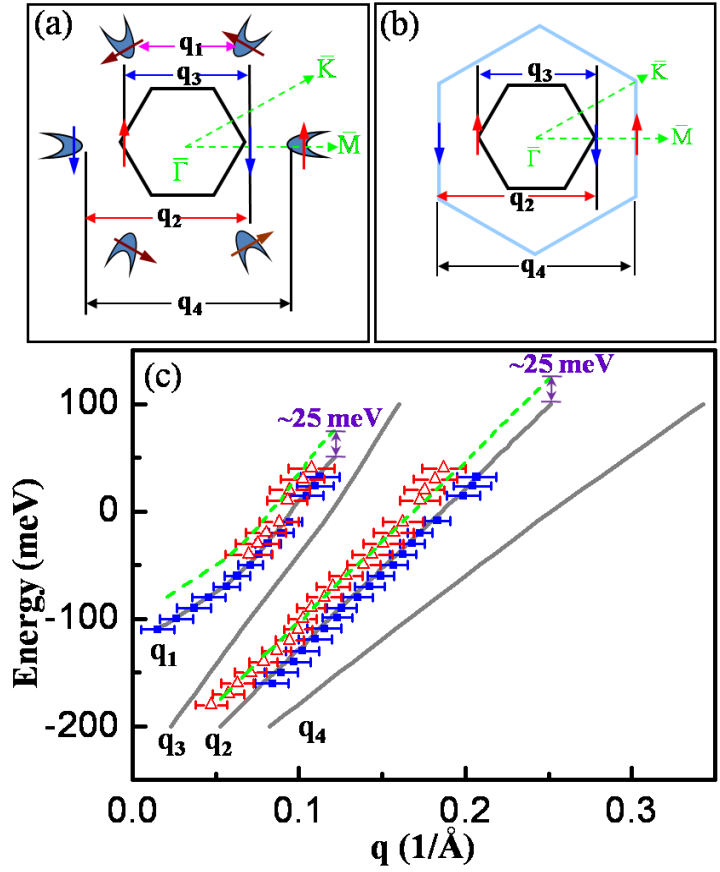

FIG. 4 Schematic of CECs at the energies (a) above -120 meV and (b) below -120 meV. All of the possible interference wavevectors ($q_1$~ $q_4$) are shown along the $\bar{\Gamma} - \bar{M}$ directions. The arrows indicate the spin orientation of the Dirac fermions. (c) Energy dispersions of the interference wavenumber. The triangular and rectangular points are obtained from the QPI patterns on the surfaces with and without Co adatoms, respectively. The solid curves correspond to the interference wavenumber ($q_1$-$q_4$) calculated from previous STM and ARPES data. The dashed curves correspond to the shifted $q_1$ and $q_2$ with the energy shifting upwards by 25 meV, respectively.